\documentclass[12pt,prl,preprint,showpacs,showkeys]{revtex4-1}
\usepackage{amsfonts}

\usepackage{graphicx}
\usepackage{dcolumn}
\usepackage{bm}
\include{biblio}

\begin{document}

\title{Plasmonic amplification and suppression in nanowaveguide coupled to gain-assisted high-quality plasmon resonances}
\author{Song-Jin Im, Gum-Song Ho}
\affiliation{Department of Physics, \textbf{Kim Il Sung} University, Daesong District, Pyongyang, DPR Korea}

\begin{abstract}
We theoretically study transmission in nanowaveguide coupled to high-quality plasmon resonances for which the metal loss is overcompensated by gain. The on-resonance transmission can vary widely from lower than --20dB to higher than 20dB for a range of gain coefficient. A reversible transition between the high-quality amplification and the suppression can be induced by a quite small change of gain coefficient for a moderately increased distance between the waveguide and the resonator. It is expected that in practice a small change of gain coefficient can be made by flexibly controlling pumping rate or utilizing nonlinear gain. Additionally, based on the frequency-dependant model for gain-transition susceptibility, it is shown that the wide variation of the on-resonance transmission can also be observed for defferent detuning of the gain-transition line-center. Such a widely controllable on-resonance transmission is promising for applications such as well-controlled lumped amplification of surface plasmon-polariton as well as plasmonic switching.
\end{abstract}

\maketitle

\section{1.	Introduction}

Plasmonic devices provide a means of guiding and manipulating light at deep subwavelength scales \cite%
{Maier_springer_2007,Barnes_nature_2003,Gramotnev_naturep_2010} and have found applications such as spectroscopy \cite%
{Stiles_rev_2008}, imaging \cite%
{Kawata_naturep_2009}, biosensing \cite%
{Anker_naturem_2008,Homola_rev_2008}, circuitry \cite%
{Ebbesen_phys_2008}, cancer treatment \cite%
{Lal_res_2008,Huang_advm_2009} and solar energy conversion \cite%
{Atwater_naturem_2010}. Surface plasmons are fundamentally lossy due to the metal losses such as the ohmic losses and the interband transitions in the metals limiting their practical applications. The use of gain media where the population inversion is created optically or electrically has been suggested for compensating the metal loss, surface plasmon amplification or spasing \cite%
{Bergman_phys_2003,Stockman_phys_2011} and experimentally demonstrated\cite%
{Oulton_nature_2009,Hill_naturep_2007,Noginov_nature_2009,Seidel_phys_2005,Exter_optxp_2013}. Stimulated emission of surface plasmon polaritons has been theoretically \cite%
{Noginov_phys_2008} and experimentally \cite%
{Leon_naturep_2010,Leon_sci_2011} studied.

In Ref. \cite%
{Yu_aphys_2008}, they proposed gain-assisted plasmonic switch consisting of a gold-air-gold plasmonic waveguide coupled to a resonator filled with gain medium for gain coefficient exactly compensating for the metal loss.

In this paper we study transmission in such a similar structure for gain coefficient larger than the exactly metal-loss-compensating one. The transmission is widely changeable from smaller than a hundredth to larger than a hundred for a range of gain coefficient larger than the exactly metal-loss-compensating one. A reversible transition between the high-quality amplification and the suppression can be induced by a quite small change of gain coefficient. Such a transition can also be induced by changing a detuning of the gain-transition frequency. For calculations we introduce a model for frequency-dependant gain coefficient.

\section{2.	Gain-assisted high-quality plasmon resonance}

Quality factor of plasmon resonance can be expressed by \cite%
{Bergman_phys_2003}
 \begin{eqnarray}
	Q=\frac{\omega\frac{\partial \mathrm{Re}s\left(\omega\right)}{\partial \omega}}{2\mathrm{Im}s\left(\omega\right)}, 
\label{eq:1}
\end{eqnarray}
 \begin{eqnarray}
	s\left(\omega\right)=\frac{\varepsilon_{h}}{\varepsilon_{h}-\varepsilon_{m}\left(\omega\right)},
\label{eq:2}
\end{eqnarray}
where $s\left(\omega\right)$ is the Bergman's spectral parameter \cite%
{Bergman_academic_1992}, $\varepsilon_{m}\left(\omega\right)$ is the permittivity of embedded metal and $\varepsilon_{h}\left(\omega\right)$ is the permittivity of embedding dielectric host medium. If we introduce a gain material in the host medium,
\begin{eqnarray}
	\varepsilon_{h}=\varepsilon_{d}+\chi_{g}
\label{eq:3}
\end{eqnarray}
where $\varepsilon_{d}$ is the permittivity of embedding dielectric excluding the gain transition contribution and $\chi_{g}$ is the gain-transition susceptibility. 

\begin{figure}
\includegraphics[width=0.8\textwidth]{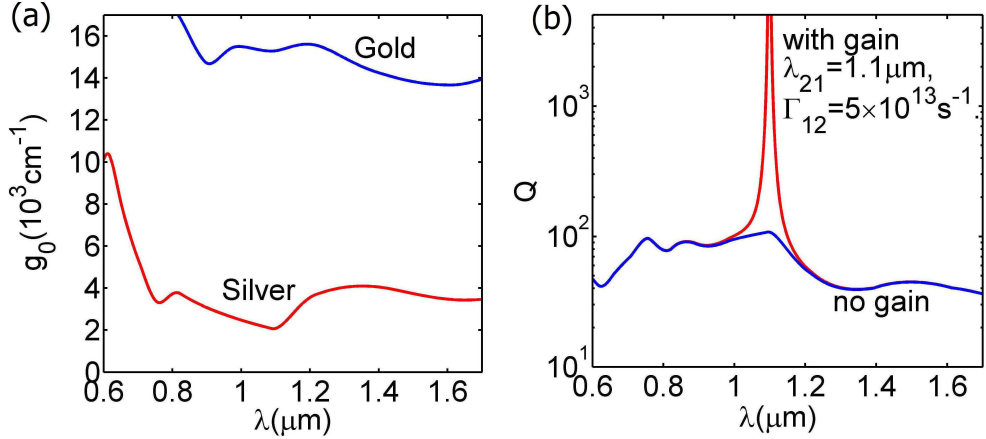}
\caption{(Color online) (a) The gain coefficients for high-quality plasmon resonances calculated by Eq. (\ref{eq:6}) for silver and gold. Here, $\varepsilon_{d}=12.25$. (b) The quality factor of plasmon resonance calculated by Eq. (\ref{eq:1}), where $\varepsilon_{m}$ is of silver. For the gain medium, $\varepsilon_{d}$ is the same as in Fig. \ref{fig:1}(a), the gain-transition line-center in terms of the wavelength $\lambda_{21}=1.1\mu$m, the dipole dephasing rate $\Gamma_{12}=5\times10^{13}\mathrm{s}^{-1}$ and the line-center gain coefficient $g_{21}=2000\mathrm{cm}^{-1}$. It is assumed that the electric field is sufficiently weak and $\Gamma^{'}=\Gamma_{12}$.}
\label{fig:1}
\end{figure}

If we reasonably assume 
\begin{eqnarray}
	\left|\varepsilon_{d}\right|>>\left|\chi_{g}\right|,\left|\mathrm{Re}\varepsilon_{m}\left(\omega\right)\right|>>\left|				\mathrm{Im}\varepsilon_{m}\left(\omega\right)\right|,
\label{eq:4}
\end{eqnarray} 
the imaginary part of the Bergman's parameter is approximately expressed as 
\begin{eqnarray}
	\mathrm{Im}s\approx\frac{\varepsilon_{d}\left|\mathrm{Re}\varepsilon_{m}\left(\omega\right)\right|}{\left[\varepsilon_{d}+\left|\mathrm{Re}\varepsilon_{m}\left(\omega\right)\right|\right]^{2}}\left[\frac{\mathrm{Im}\varepsilon_{m}\left(\omega\right)}{\left|\mathrm{Re}\varepsilon_{m}\left(\omega\right)\right|}-\frac{\left|\mathrm{Im}\left[\chi_{g}\right]\right|}{\varepsilon_{d}}\right].
\label{eq:5}
\end{eqnarray}   
From Eq. (\ref{eq:1}) and (\ref{eq:5}), we can obtain the gain coefficients for high-quality plasmon resonances       		\begin{eqnarray}
		g_{0}\left(\omega\right)=\frac{\sqrt{\varepsilon_{d}}\omega}{c}\frac{\mathrm{Im}\varepsilon_{m}\left(\omega\right)}{\left|\mathrm{Re}\varepsilon_{m}\left(\omega\right)\right|}.
\label{eq:6}
\end{eqnarray}
In the derivation of Eq. (\ref{eq:6}), we considered that the gain coefficient is given in general by $g=2\left|\mathrm{Im}\sqrt{\varepsilon}\right|\omega/c=\omega\left|\mathrm{Im}\varepsilon\right|/c\sqrt{\mathrm{Re}\varepsilon}\approx\omega\left|\mathrm{Im}\chi_{g}\right|/c\sqrt{\varepsilon_{d}}$.
Eq. (\ref{eq:6}) is essentially same as the spasing criterion $g_{th}$ in Ref. \cite%
{Stockman_optxp_2011} and the condition for large quality factor in Ref. \cite%
{Wang_phys_2006} under the assumption (\ref{eq:4}). Fig. \ref{fig:1}(a) shows $g_{0}$ according to wavelength for silver and gold assuming the dielectric permittivity $\varepsilon_{d}=12.25$. The experimental data for the permittivity of slilver and gold according to wavelength \cite%
{Johnson_phys_1972} are directly used. As shown in Fig. \ref{fig:1}(a), silver is better than gold for high-quality plasmon resonance. Moreover, resonance wavelengths near 1.1$\mu$m for silver are promising if we consider that a gain coefficient greater than $3000\mathrm{cm}^{-1}$is technically difficult to realize \cite%
{Stockman_optxp_2011}.

We take account of the frequency dependency of the gain-transition susceptibility $\chi_{g}$. The frequency-dependant gain-transition susceptibility $\chi_{g}\left(\omega\right)$  can be expressed in the two-level approximation \cite%
{Boyd_pergamon_2008} by
\begin{eqnarray}
		\chi_{g}\left(\omega\right)=g_{21}\frac{c\sqrt{\varepsilon_{d}}}{\omega_{21}}\frac{\left(\omega-\omega_{21}\right)/\Gamma_{12}-i}{1+\left(\omega-\omega_{21}\right)^2/\Gamma^{'^{2}}},
\label{eq:7}
\end{eqnarray}
where $\omega_{21}$ is the gain-transition line-center, $\Gamma_{12}$ is the dipole dephasing rate, $g_{21}$  is the line-center gain coefficient and $\Gamma^{'}$ is the gain-transition line width. For a sufficiently strong electric field \cite%
{Boyd_pergamon_2008}, 
\begin{eqnarray}
		g_{21}=\frac{g_{21}^{0}}{1+\left|E\right|^{2}/\left|E_{0}^{s}\right|^{2}},  \
		\Gamma^{'}=\Gamma_{12}\left(1+\left|E\right|^{2}/\left|E_{0}^{s}\right|^{2}\right)^{1/2},
\label{eq:8}
\end{eqnarray}
where $E_{0}^{s}$ is the line-center saturation field strength and $g_{21}^{0}=\omega_{21}N\left(\rho_{bb}-\rho_{aa}\right)^{(eq)}\left|\mu_{21}\right|^{2}/\left(\sqrt{\varepsilon_{d}}\Gamma_{12}\varepsilon_{0}c\eta\right)$ \cite%
{Boyd_pergamon_2008} is the line-center gain coefficient for a weak electric field compared to $E_{0}^{s}$ .

Fig. \ref{fig:1}(b) shows gain-assisted improvement of plasmon resonance quality. The quality factor is calculated by Eq. (\ref{eq:1}). The embedded metal is silver. For the gain medium, $\varepsilon_{d}$ is the same as in Fig. \ref{fig:1}(a), the gain-transition line-center in terms of the wavelength $\lambda_{21}=1.1\mu$m, the dipole dephasing rate $\Gamma_{12}=5\times10^{13}\mathrm{s}^{-1}$ \cite%
{Stockman_optxp_2011} and the line-center gain coefficient $g_{21}=2000\mathrm{cm}^{-1}$ corresponding to $g_{0}\left(\lambda_{21}\right)$ . It is assumed that the electric field is sufficiently weak and $\Gamma^{'}=\Gamma_{12}$. As expected, in the narrow spectral range around 1.1$\mu$m a peak of quality factor appears.          

\begin{figure}
\includegraphics[width=0.6\textwidth]{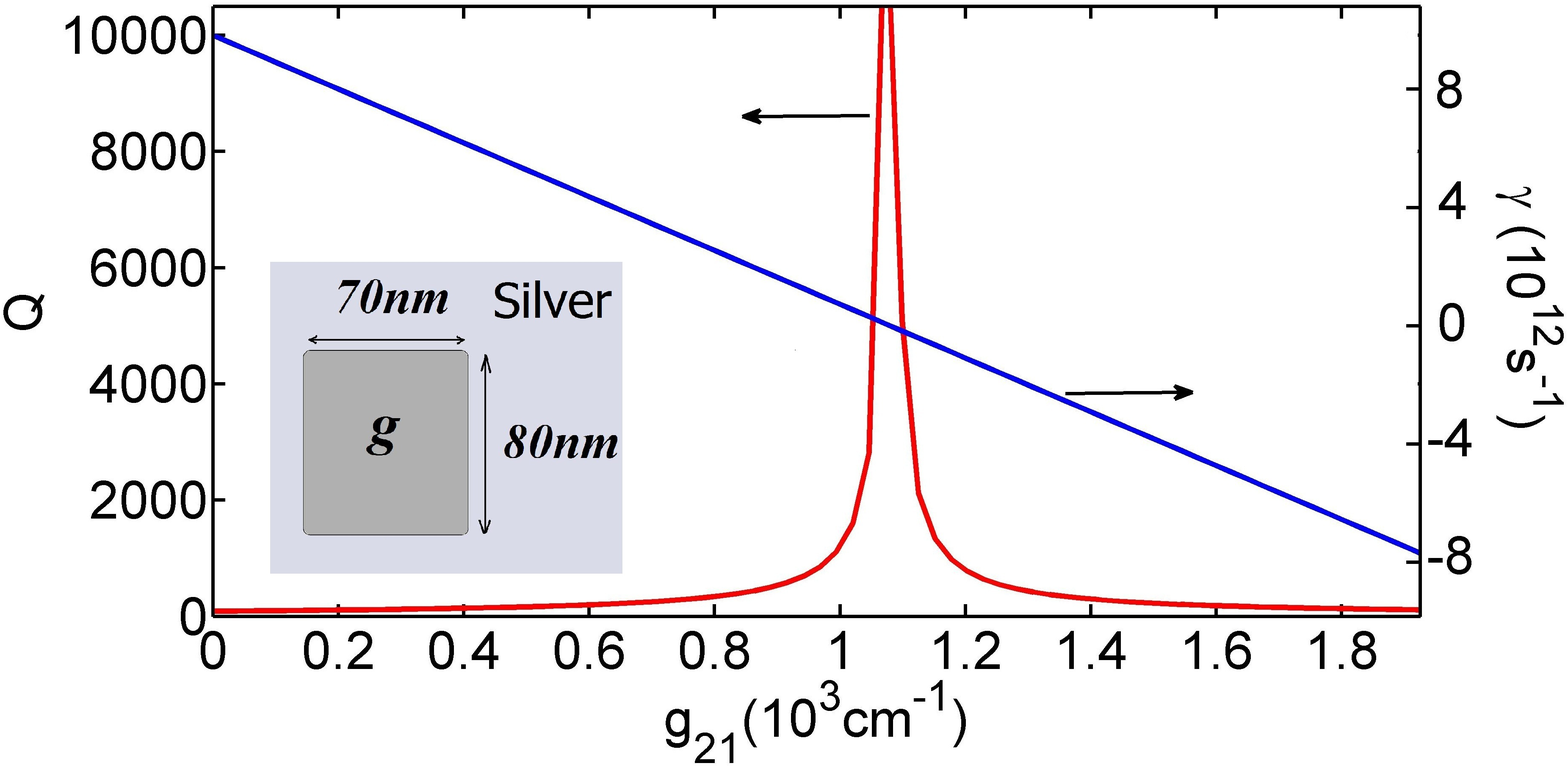}
\caption{(Color online) The quality factor Q and relaxation rate $\gamma$ in two-dimensional rectangular nanostructure with dimensions of 70nm and 80nm filled with gain medium embedded in silver (shown in the insect) according to the line-center gain coefficient $g_{21}$. For the gain medium, $\lambda_{21}=\lambda_{res}=1.09\mathrm{\mu m}$ and other parameters are the same as in Fig. \ref{fig:1}(b).}
\label{fig:2}
\end{figure}

We choose a simple two-dimensional rectangular nanostructute with dimensions of 70nm and 80nm filled with dielectric with the permittivity $\varepsilon_{d}=12.25$ embedded in silver (shown in the insect of Fig. \ref{fig:2}) as a plasmonic resonator with resonance wavelength $\lambda_{res}=1.1\mathrm{\mu m}$. When solving directly the Maxwell equation in frequency domain as an eigenfrequency problem taking account of retardation effects, the eigenfrequency is $\omega_{res}-i\gamma$, where $\omega_{res}$ and $\gamma$ means a resonance frequency and a relaxation rate due to the loss in metal, respectively. For the above nanostructure, the numerical calculation shows $\omega_{res}=1.73\times10^{15}\mathrm{s}^{-1}\left(\lambda_{res}=1.09\mathrm{\mu m}\right)$, $\gamma=9.8\times10^{12}\mathrm{s}^{-1}$ and $Q=\omega_{res}/2\left|\gamma\right|$=88. The quality factor Q of the plasmon resonance can be increased by introducing a gain material to the dielectric rectangle as shown in Fig. \ref{fig:2}. For the gain material, $\lambda_{21}=\lambda_{res}$ and $\Gamma_{12}$ is the same as in Fig. \ref{fig:1}(b). The line-center gain coefficient $g_{21}$=1080$\mathrm{cm}^{-1}$ for the maximum Q and the zero $\gamma$, where the metal loss is exactly compensated, is significantly smaller than the $g_{0}\left(\lambda_{res}\right)=2080\mathrm{cm}^{-1}$ expected in Fig. \ref{fig:1}(a). This discrepancy is attributed to retardation effects which is ignored in the quasistatic approximation in the derivation of Eq. (\ref{eq:1}). Near the maximum Q, a sign of relaxation rate $\gamma$ is changed. For a positive relaxation rate $\gamma$ the metal loss relaxes the local field to a stationary state, while the negative sign mathematically means that the local field is exponentially enhanced for $t\rightarrow+\infty$ by a gain overcompensating the metal loss, which is physically impossible. In fact, Eq. (\ref{eq:8}) shows an sufficiently enhanced field decreases the gain coefficient so that the local field goes to a stationary state with a great quality factor Q. Thus, for a line-center gain coefficient $g_{21}$ greater than the criterion value of 1080$\mathrm{cm}^{-1}$ , the rectangular nanoresonator can act as a two-dimensional spaser. 

\section{3.	Amplification and suppression in nanowaveguide coupled to gain-assisted high-quality plasmon resonances}

\begin{figure}
\includegraphics[width=0.8\textwidth]{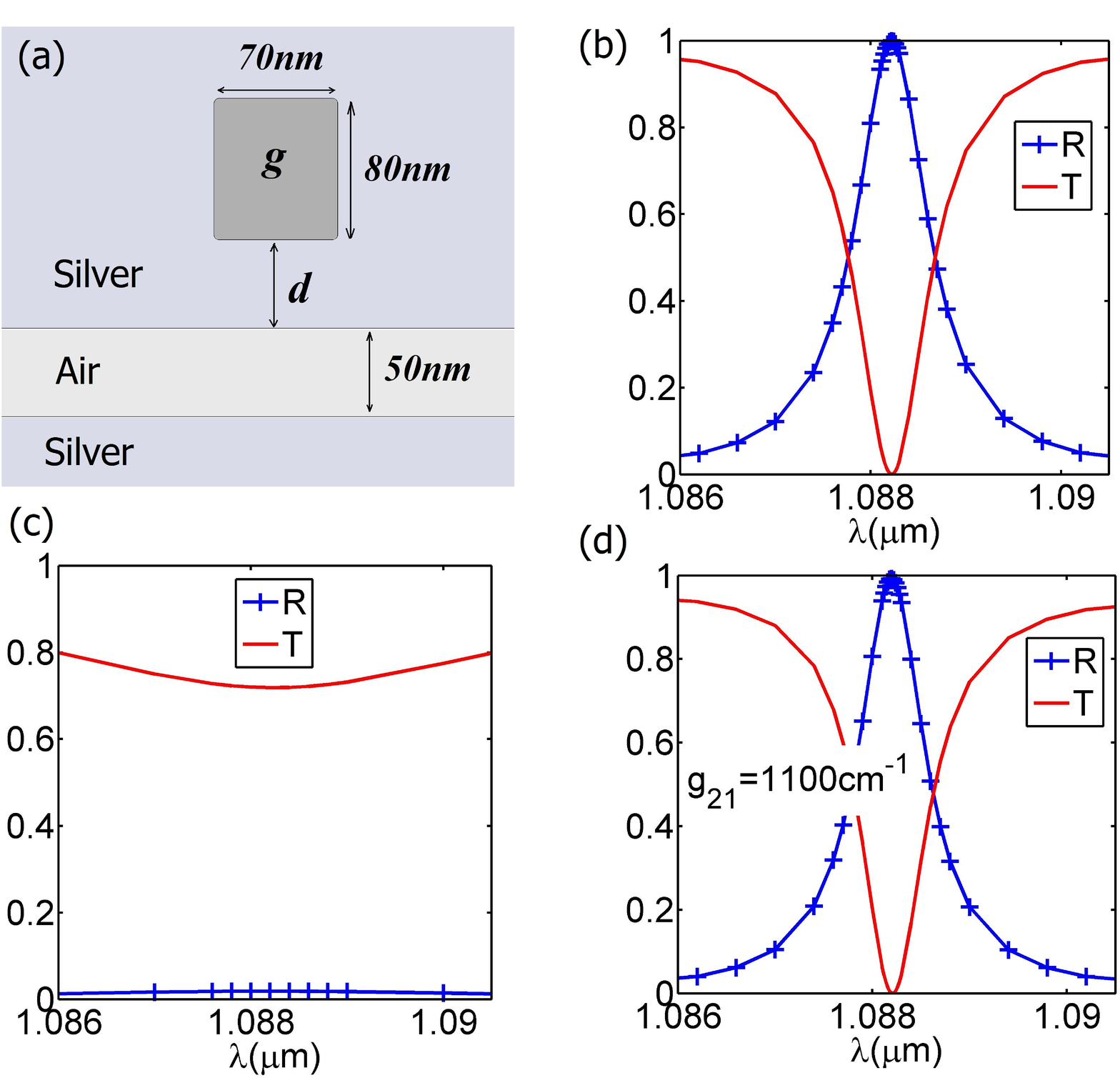}
\caption{(Color online) (a) Transmission T and reflection R spectra in a silver-air-silver nanowaveguide side coupled to the plasmonic resonator [shown in the insect of (a)] for $d=50\mathrm{nm}$. (b) the metal-loss-free state neglecting the imaginary part of the permittivity of silver. (c) the full-loss state considering the metal loss. (d) same as (c) except that the gain $g_{21}=1100\mathrm{cm}^{-1}$ is included. For the gain medium, other parameters are the same as in Fig. \ref{fig:2}.}
\label{fig:3}
\end{figure}

We calculate transmission T and reflection R coefficients in a metal-insulator-metal (MIM) nanowaveguide side coupled to the above gain-assisted plasmonic resonator [shown in Fig. \ref{fig:3}(a)] by numerically solving the Maxwell equation in frequency domain. Our calculations well agree with the coupled-mode theory \cite%
{Yu_aphys_2008,Haus_ieee_1992}.

In Ref. \cite%
{Yu_aphys_2008} a quite similar structure was studied mainly focusing on the range between the high-loss state without pumping and the exact metal-loss-compensation state with pumping. Fig. \ref{fig:3}(b)-(d) shows essentially the same facts as in Ref. \cite%
{Yu_aphys_2008} that the metal-loss-free state [shown in Fig. \ref{fig:3}(b)] neglecting the imaginary part of the permittivity of silver can be reproduced for a certain gain coefficient [shown in Fig. \ref{fig:3}(d)], while Fig. \ref{fig:3}(c) shows the full-loss state without gain. The $g_{21}=1100\mathrm{cm}^{-1}$ [shown in Fig. \ref{fig:3}(d)] almostly coincidences with the $g_{21}=1080\mathrm{cm}^{-1}$ for the exact metal-loss-compensation [shown in Fig. \ref{fig:2}]. The difference between these two gain coefficients is attributed to a metal loss in the MIM waveguide. 

\begin{figure}
\includegraphics[width=0.8\textwidth]{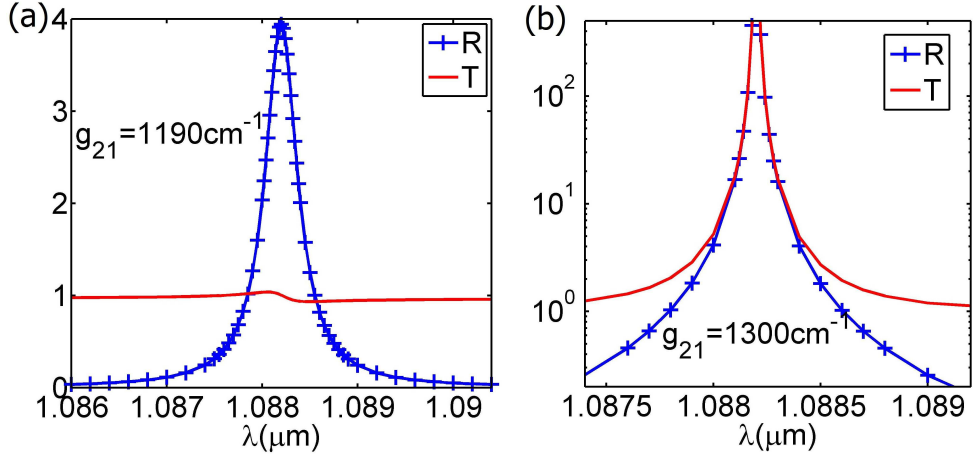}
\caption{(Color online) Transmission T and reflection R spectra in the nanowaveguide side coupled to the plasmonic resonator same as in Fig. \ref{fig:3} with a gain coefficient larger than the one for the exact metal-loss-compensation. (a) Broadband full transmission T=1 and on-resonance amplified reflection $R\left(\lambda_{res}\right)=4$ for $g_{21}=1190\mathrm{cm}^{-1}$. (b) High-quality amplification of both transmission and reflection for $g_{21}=1300\mathrm{cm}^{-1}$. For the gain medium, other parameters are the same as in Fig. \ref{fig:2}.}
\label{fig:4}
\end{figure}

\begin{figure}
\includegraphics[width=0.6\textwidth]{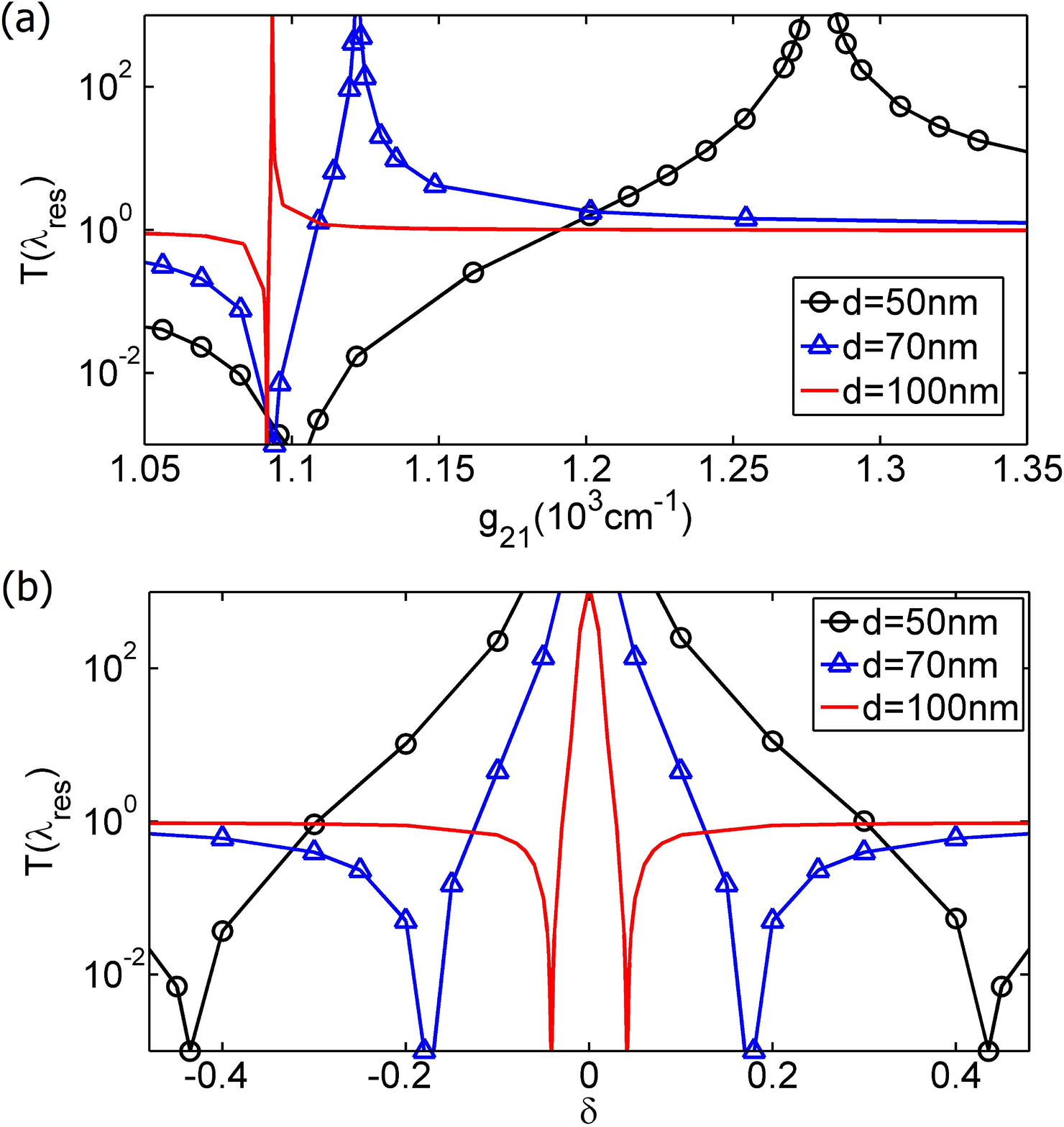}
\caption{(Color online) On-resonance transmission of the waveguide coupled to the gain-assisted plasmonic resonator [shown as Fig. \ref{fig:3}(a)] with different distances $d$ according to the line-center gain coefficient $g_{21}$ (a) and the normalized detuning of the gain-transition line-center $\delta=\left(\omega_{21}-\omega_{res}\right)/\Gamma_{12}$(b). In (b) $g_{21}=1300\mathrm{cm}^{-1}$ for $d=50\mathrm{nm}$, $g_{21}=1120\mathrm{cm}^{-1}$ for $d=70\mathrm{nm}$ and $g_{21}=1090\mathrm{cm}^{-1}$ for $d=100\mathrm{nm}$ corresponding to the high-quality amplification peaks in (a). Other parameters for the gain medium are the same as in Fig. \ref{fig:2}. }
\label{fig:5}
\end{figure}

Whereas the rectangular plasmonic resonator with a $g_{21}$ larger than the exactly metal-loss-compensating one has no physically stable state as shown in Fig.  \ref{fig:2}, the resonator with such a larger $g_{21}$ coupled to waveguide has stable states as shown in Fig. \ref{fig:4}, where an excess energy produced by overwhelming gain can leak from the resonator through the waveguide. For $g_{21}=1190\mathrm{cm}^{-1}$, broadband full transmission T=1 and on-resonance amplified reflection $R\left(\lambda_{res}\right)=4$ can be seen in Fig. \ref{fig:4}(a). A further larger gain coefficient can give rise to a high-quality amplification of both transmission and reflection as shown in Fig. \ref{fig:4}(b).

Transition between the suppression [Fig. \ref{fig:3}(d)] with the zero of on-resonance T and the high-quality amplification [Fig. \ref{fig:4}(b)] with a large on-resonance T can be induced by a change of gain coefficient as shown in Fig.  \ref{fig:5}(a). The interval between the gain coefficients for the high-quality amplification and the suppression rapidly decreases with the increase of the distance $d$ between the resonator and the waveguide and can be quite small for a moderately increased $d$. The longer distance $d$ results in the weaker coupling between the waveguide and the resonator and thus the less leakage from the resonator which can be compensated by the smaller increment of gain coefficient. Here, it is noted that the position of gain coefficient for the suppression is independent on the distance $d$. The transition can be also induced by changing the normalized detuning of the gain-transition line-center $\delta=\left(\omega_{21}-\omega_{res}\right)/\Gamma_{12}$ as can be seen in Fig. \ref{fig:5}(b).

\section{4.	Conclusions}

The transmission in nanowaveguide coupled to high-quality plasmon resonances for which the metal loss is overcompensated by gain can vary in wide range from lower than --20dB to higher than 20dB for a range of gain coefficient, width of which can be quite small for a moderately increased distance between the waveguide and the resonator. We expect that in practice a small change of gain coefficient can be made by flexibly controlling pumping rate or utilizing nonlinear gain. The widely variable transmission can also be observed for defferent detuning of the gain-transition line-center. The widely controllable on-resonance transmission is promising for applications such as plasmonic switching and lumped amplification of surface plasmon-polariton.

\end{document}